\documentclass[prb,twocolumn,showkeys,superscriptaddress]{revtex4}
\usepackage{graphicx}
\usepackage{amsmath}
\usepackage[colorlinks=true,linkcolor=blue,urlcolor=blue,citecolor=blue]{hyperref}
\newcommand*{\Vshift}{V_{\mathrm{shift}}}
\newcommand*{\Vgad}{V_{\mathrm{g,ad}}}
\newcommand*{\Vcnp}{V_{\mathrm{CNP}}}
\newcommand*{\Nox}{N_{\mathrm{ox}}}
\newcommand*{\nimp}{\overline{n_{\mathrm{ox}}}}
\newcommand*{\diff}[2]{\frac{\mathrm{d} #1}{\mathrm{d} #2}}
\newcommand*{\tdiff}[2]{\mathrm{d} #1/\mathrm{d} #2}
\newcommand*{\feref}{\varepsilon _\mathrm{F}}
\newcommand*{\kbt}{k_\mathrm{B}T}
\newcommand*{\Pkin}{P kinetics}
\newcommand*{\Hkin}{H kinetics}
\newcommand{\titech}{Department of Chemistry, Tokyo Institute of Technology, 2-12-1 Ookayama, \\
Meguro-ku, Tokyo, 152-8551, Japan}
\begin{document}

\affiliation{\titech}
\author{Yoshiaki Sato}
\email{mail to: sato.y.an@m.titech.ac.jp}

% \fax{+81-3-5734-2242}
% \phone{+81-3-5734-2242}

\author{Kazuyuki Takai}
\affiliation{\titech}
\author{Toshiaki Enoki}
\affiliation{\titech}

\title{Electrically Controlled Adsorption of Oxygen in Bilayer Graphene Devices}
\keywords{Graphene; charge transfer; field effect transistor; 
electron transport; mobility; band gap}

% \begin{document}

\begin{abstract}

We investigate the chemisorptions of oxygen molecules on bilayer graphene (BLG)
and its electrically modified charge-doping effect
using conductivity measurement of the field effect transistor channeled with BLG. 
We demonstrate that the change of the Fermi level
by manipulating the gate electric field significantly affects not only the rate of molecular adsorption
but also the carrier-scattering strength of adsorbed molecules.
Exploration of the charge transfer kinetics reveals the electrochemical nature of the oxygen adsorption on BLG.
[This document is the unedited Author's version of 
a Submitted Work that was subsequently accepted for 
publication in \textit{Nano Letters}, \copyright American
Chemical Society after peer review. To access the final edited 
and published work see \url{http://dx.doi.org/10.1021/nl202002p}.]
\end{abstract}

\maketitle

It has been a central topic of surface science how to control the 
adsorption and desorption in order to to bring out desirable features 
and functionalities by adsorbed molecules. 
Tuning the electronic features of solid surfaces has an important implication in that 
molecular chemisorptions and catalytic reactions are determined by them\cite{BalogR2010,ZhangY2004}.
In particular for graphene, the two-dimensional honeycomb 
carbon lattice, in which the conduction $\pi ^{*}$-band and the 
valence $\pi $-band contact to each other at the ``Dirac point'' 
giving a feature of zero-gap semiconductor\cite{CastroNetoAH2009}, the control of 
chemisorption is a critical issue since chemisorption directly leads to 
altering every electronic property of graphene. Other than the 
electron/hole doping\cite{SchedinF2007} owing to the charge transfer between 
graphene and the adsorbed molecules, widely known are the charged impurity 
effect on the electron transport\cite{ChenJH2007, XiaoS2009, ChenJH2009},
lattice deformation\cite{BoukhvalovDW2009}, and opening the band gap
due to asymmetric adsorption\cite{CastroEV2010a, AbaninDA2010, RibeiroRM2008}.
Aside from the macroscopic spatially-controlled adsorption that is achieved using 
nano-device fabrication technique\cite{LohmannT2009}, microscopic control of 
adsorption structure is of great importance because the aforementioned 
adsorption effects are altered by the local structure of adsorbate, e.g., 
whether the adsorbed molecules are arranged in a random or superlattice
structure\cite{AbaninDA2010, CheianovVV2010}, 
or whether the molecules are adsorbed individually\cite{SchedinF2007, RomeroHE2009} 
or collectively (in dimers\cite{WehlingTO2009a} or
clusters\cite{TakaiK2007, McCrearyKM2010}). For the 
first step to realize such an advanced control of adsorption, the 
methods to utilize the interaction between the adsorbed molecules and 
graphene for it are to be explored.

The principal impetus in the present study is to control the charge 
transfer between graphene and the adsorbed molecules by tuning the Fermi 
level of graphene, which is readily accomplished in the field effect 
transistor (FET) structure. When $\mathrm{SiO_2}$/Si substrate is used as 
the back-gate insulator of the FET, the tuning range of the Fermi level 
of graphene by the application of the gate voltage is at the extent of 
several $\pm 0.1$~eV\cite{NovoselovKS2005} which would be sufficient to alter the 
chemical reactivity on the surface. Besides, the additional charge and 
the gradient of electric potential generated by the gate electric field 
are expected to change the polarization of adsorbed
molecules\cite{ZhouJ2010, SaalfrankP2000} and to modify the charge
distribution on graphene layers and the adsorbed
molecules\cite{LiuW2009, LuYH2009a, BrarVW2010, ParqJH2010, TianX2010}
leading to, e.g., the change in the ease of migration of molecules
adsorbed on graphene\cite{SuarezAM2011}. 
Some research\cite{KaverzinAA2011, PintoH2010, YangY2011}
argues that the change in the Fermi level 
caused by a gate electric field activates electrochemical redox 
reactions and the accompanying charge transfer causes hysteresis of the 
source--drain current in graphene FET, yet there has been no 
investigation that elucidates the relation between the kinetics of 
adsorption to graphene and gate electric field. In this study we 
investigated gate-tuned molecular oxygen adsorption through systematic 
measurements of conductivity using mainly bilayer graphene (BLG).

\begin{figure}
	\centering
	\includegraphics[width=\linewidth, clip]{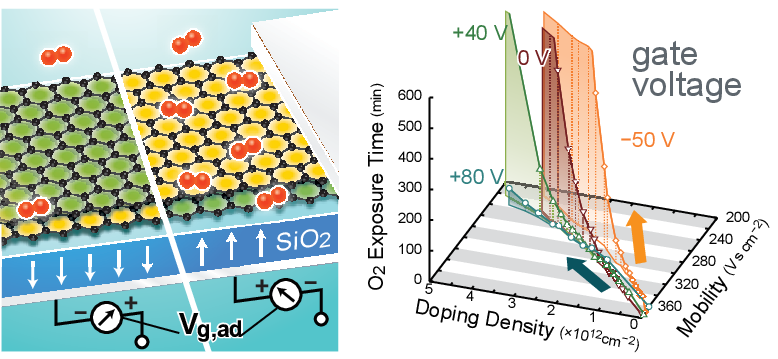}
\end{figure}

\begin{figure*}
	\centering
	\includegraphics[clip]{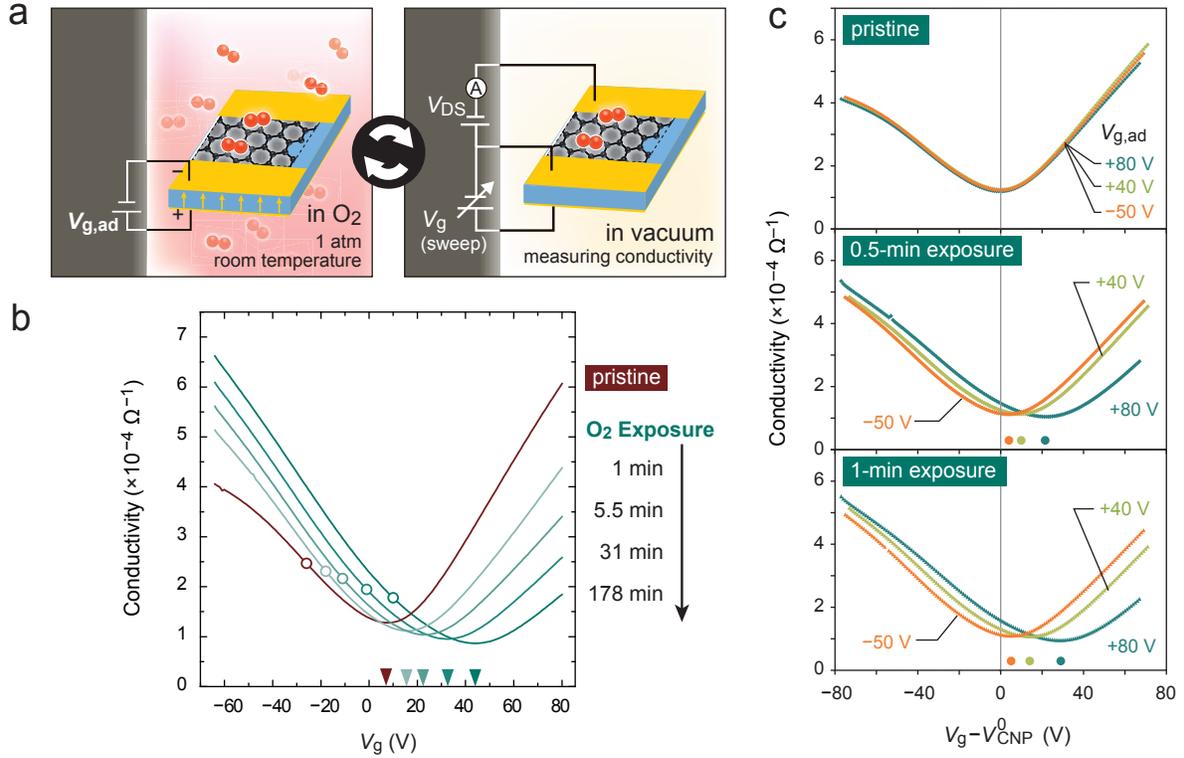}
	\caption{(a) Schematic of the measurement cycle. First, the field effect 
	transistor channeled by bilayer graphene (BLG-FET) is exposed to gaseous 
	$\mathrm{O_2}$ while the gate voltage $\Vgad $ is applied (left panel). 
	Then the system is evacuated and the source--drain conductivity of the 
	BLG-FET is measured by sweeping the gate voltage $V_\mathrm{g}$ (right panel). 
	Subsequently gaseous $\mathrm{O_2}$ is again introduced, and the cycle is 
	repeated. The gas introduction and evacuation are completed in 
	a shorter time than $\sim 10$ s to prevent additional gas from adsorption. 
	The whole cycle is executed at room temperature. (b) Change of the 
	field effect behavior due to the $\mathrm{O_2}$ exposure with $\Vgad =0$~V
	(run~1). The gate voltage 
	giving the minimum conductivity (the charge neutrality point), 
	is shifted from $V_\mathrm{g}=\Vcnp^{0}$ ($<8$~V) 
	(marked by a brown triangle, before $\mathrm{O_2}$ exposure) 
	to the positive direction (green triangles) upon $\mathrm{O_2}$ exposure.
	 Circles on the curves represent the conductivity at the hole 
	density of $2.5\times 10^{12}\ \mathrm{cm^{-2}}$ 
	that are used to calculate Drude conductivity shown in Figure~\ref{f:mobility}a. 
	(c) The same measurement as in the panel (b) with applying the finite $\Vgad $
	. All the curves are shifted by $-\Vcnp^{0}$
	($\Vcnp^{0} = 13$, 9, and 11~V for the run of $\Vgad =+80$, $+40$,
	and $-50$~V, respectively)
	in $V_\mathrm{g}$ direction, 
	i.e., the charge neutrality points of the pristine graphene 
	without the adsorbed oxygen are taken as 
	zero gate voltage. The top panel of (c) represents the $\sigma $~vs 
	$V_\mathrm{g}-\Vcnp^{0}$ for the pristine graphene.
	The changes of $\sigma$~vs $V_\mathrm{g}-\Vcnp^{0}$ curve after a single and a double 
	exposure to $\mathrm{O_2}$ (the time duration of a single exposure is 30 
	s) are shown in the center and the lower panel of (c), respectively. 
	Filled circles indicate $\Vshift=\Vcnp-\Vcnp^{0}$ (the shift of 
	the CNP) for each curve.}%
	\label{f:intro}
\end{figure*}

The back-gated BLG-FETs were fabricated on $\mathrm{SiO_2}$ (300~nm thick) 
on heavily n-doped silicon substrate by means of 
photolithography. The channel length and width were 6~$\mathrm{\mu m}$ and 3.5~$\mathrm{\mu m}$, 
respectively. Prior to measurement, we repeated vacuum annealing ($210^\circ\mathrm{C}$, 
10~h) to remove the adsorbed moisture and contaminants on the surface 
until no more changes in the gate-dependent conductivity $\sigma $ were 
eventually seen. After the annealing, the BLG-FET exhibited its pristine 
nature, that is, ambipolar transport properties with a conductivity 
minimum around $V_\mathrm{g}=\Vcnp^{0}<8$~V giving the ``\,charge neutrality 
point (CNP)'' with electrons and holes in BLG being equal in density. The Drude 
mobility $\mu (n)$ was estimated to be $\sim 1\times 10^{3}\ \mathrm{cm^{2} V^{-1} 
{s}^{-1}}$ from the equation $\mu(n) =\sigma /e\vert n\vert $, where $n$ is 
the carrier density with $n=({c_\mathrm{g}}/{e})(V_\mathrm{g}-\Vcnp)$ ($
c_\mathrm{g}/e=7\times 10^{10}\ \mathrm{cm^{-2}V^{-1}}$; $c_\mathrm{g}$ is the 
capacitance per unit area for the back-gated graphene FET on 300 
nm-thick $\mathrm{SiO_2}$ )\cite{LuYH2009a}, It is in the range of the values for 
BLG-FET in two-probe configuration previously reported\cite{XiaoS2009}, and 
therefore we confirm that the graphene of the present BLG-FET has few 
defects that may extremely enhance the chemical reactivity of graphene
\cite{deAndressPL2008}. 

\begin{figure*}
	\centering
	\includegraphics[clip,scale=1.25]{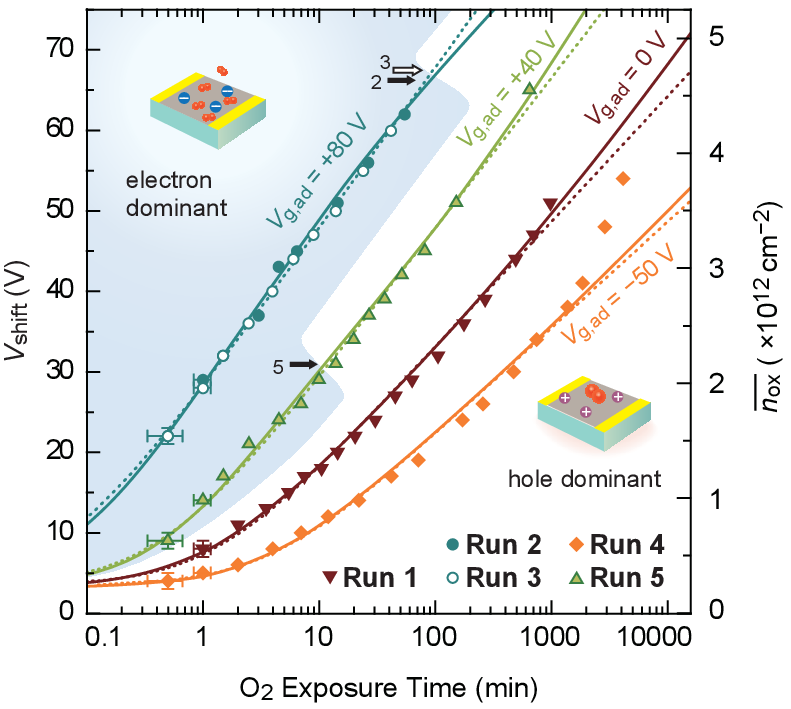}
	\caption{
	Time dependence of $\Vshift$ and doping density $\nimp$ 
	with doping due to the $\mathrm{O_2}$  exposure under application of various 
	$\Vgad $. Solid and dotted curves are the fits based on the 
	\Hkin\ and the \Pkin\ (see text), respectively. The curve fitting is 
	made in the range of exposure time below 2000~min. The electron-dominant 
	region, where the Fermi level of BLG is higher than the CNP 
	($\Vgad>\Vcnp$), is painted blue and the hole-dominant region 
	($\Vgad<\Vcnp$) is painted white. Carrier type is inverted
	between electron and hole at the 
	point indicated by arrows (numbers aside correspond to the run number) 
	during the evolution of the oxygen adsorption.}
	\label{f:timeevolve}
\end{figure*}

Next we exposed the BLG-FET to 1 atm of high-purity ($>$99.9995\%) 
oxygen in the measurement chamber at room temperature. Instead of 
measuring conductivity with graphene kept in the $\mathrm{O_2}$ environment, 
we performed the short-time interval exposure--evacuation cycles 
schematically shown in Figure~\ref{f:intro}a; $\mathrm{O_2}$ exposure was done under the dc 
gate voltage $\Vgad$, followed by rapid evacuation in less 
than 10~s (the physisorbed $\mathrm{O_2}$ molecules would be removed 
immediately without charge transfer), and eventually the $\sigma $~vs $
V_\mathrm{g}$ measurement was done  with sweeping  $V_\mathrm{g}$ 
under vacuum. In this cycle, we can rule out the possibility of the 
additional oxygen adsorption during sweeping gate voltage for the $ 
\sigma $~vs $V_\mathrm{g}$ measurement since the system was evacuated then. 
In addition, we found that the $\sigma $~vs $V_\mathrm{g}$ curve did 
not vary under vacuum at room temperature at least for more than several 
hours, so that we can also rule out the possibility of the oxygen 
desorption during $\sigma $~vs $V_\mathrm{g}$ measurement (taking ca. 10 min to obtain a single $\sigma $~vs $V_\mathrm{g}$ curve). 
Therefore,
just repeating the cycles substantially realizes the long-time
$\mathrm{O_2}$ exposure under $\Vgad$, 
the length of which is denoted by total $\mathrm{O_2}$ exposure time, $t$.
Figure~\ref{f:intro}b shows the change in $\sigma 
$~vs $V_\mathrm{g}$ by repeating the $\mathrm{O_2}$ exposure-evacuation 
cycles without applying gate voltage during $\mathrm{O_2}$ exposure (run~1, 
$\Vgad=0$~V).
The shift of the charge neutrality point by the amount of
$\Vshift(t)=\Vcnp(t)-\Vcnp^{0}$ toward the positive 
direction was observed (Figure~\ref{f:intro}b), which represents 
hole doping to graphene. Prolonged exposure brought further hole doping, 
and eventually the doping density (induced charge by the oxygen
adsorption) $\nimp(t)=({c_\mathrm{g}}/{e})\Vshift(t)$ reached more 
than $5\times 10^{12}\ \mathrm{cm^{-2}}$ within a time scale of $10^{3}$ min. 
Note that any hysteresis as observed in graphene FET in moist 
atmosphere\cite{ShiY2009, AguirreCM2009, LevesquePL2010, SabriSS2009, YangY2011}
was not found in the observed $\sigma $ 
vs $V_\mathrm{g}$ curve, so that we can uniquely
determine $\Vcnp(t)$ as a function of $t$.
Another remarkable feature is the hole conductivity in highly doped 
regime, $V_\mathrm{g}-\Vcnp(t)<-40$~V (i.e., $\vert n\vert >3\times 10^{12}\
\mathrm{cm^{-2}}$). The $\sigma $~vs $V_\mathrm{g}$ curve distorted and exhibited 
the sublinear dependence in this regime for the pristine BLG-FET
as can be seen in Figure~\ref{f:intro}b. Yet 
it disappeared only after the exposure to $\mathrm{O_2}$ for 1 min, whereas 
the carrier doping has not proceeded much at that time. Thus this rapid 
change in conductivity feature is discriminated from the slower change 
causing the shift of the CNP; one possibility is that the former is due 
to rapid oxidation of the metal--graphene interface\cite{NouchiR2010, YamadaT2004}, which 
does not shift the Fermi level of graphene but asymmetrically varies the 
conductivity. 

Both $\Vcnp$ and the mobility of the BLG-FET with the oxygen 
adsorbed were reset to the value for the pristine BLG-FET by annealing 
the $\mathrm{O_2}$-exposed BLG-FET in vacuum at 200~$^{\circ}\textrm{C}$,
indicating that $\mathrm{O_2}$ desorption readily proceeds at  high temperature without 
making any defects. By virtue of this reversibility of the oxygen 
adsorption, we can repeat the conductivity measurements in the 
exposure--evacuation cycles as described above for the same device and 
compare the results. 
We additionally carried out four consecutive measurements under the
same condition except that a finite gate voltage $\Vgad $ was 
applied during $\mathrm{O_2}$ exposure; $\Vgad $ was $+80$ 
V (run~2 and run~3), $-50$~V (run~4) and $+40$~V (run~5). 
Figure~\ref{f:intro}c represents the change of the 
gate-dependent conductivity $\sigma $~vs $V_\mathrm{g}-\Vcnp^{0}$ at the 
initial step of the runs~2, 4, 5: before $\mathrm{O_2}$ exposure, 
(i.e., after vacuum annealing) and after 
the first and second exposure--evacuation cycles
 (the duration time for $\mathrm{O_2}$ exposure in each cycle is 30 s,
i.e., $t=0.5$, $1$ min, after the first and second cycle, respectively). All the $
\sigma $~vs $V_\mathrm{g}-\Vcnp^{0}$ curves collapsed onto almost the 
identical curve in the pristine graphene as shown in the top panel of Figure~\ref{f:intro}c. 
Since $\Vcnp^{0}$
was within $10\pm 3$~V for each run (see 
the caption of Figure~\ref{f:intro}), the BLG-FET was realized to exhibit its pristine 
feature before $\mathrm{O_2}$ exposure. After the $\mathrm{O_2}$ exposure (the 
center and the bottom panel of Figure~\ref{f:intro}c) the gate-dependent conductivity 
changes similarly as was also observed for  run~1 (Figure~\ref{f:intro}b), but the 
effect of applying $\Vgad$ during $\mathrm{O_2}$ exposure is marked by 
the clear difference in $\Vshift(t)$. 
There is a tendency that
$\Vshift(t)$ is larger for higher $\Vgad $ and smaller for lower
$\Vgad $, indicating that hole doping proceeds more intensively to 
graphene with higher Fermi level. This trend is pronounced on the 
increase in the exposure time, as confirmed by comparing the center and 
the bottom panel of Figure~\ref{f:intro}(c).

We tracked the temporal evolution of the gate-dependent conductivity 
over a wide time range between $10^{0}$--$10^{3}$ min. 
Figure~\ref{f:timeevolve} represents $\Vshift$ for runs~1--5 with 
respect to  $\mathrm{O_2}$ exposure time, in which the corresponding doping 
density owing to the oxygen adsorption, $\nimp$, is also shown in the right axis. 
The 
tendency that the high $\Vgad$ leads to rapid doping can be seen 
clearly over a whole time range; e.g., to reach the doping level of
$\Vshift=40$~V, it took ca.~300 min for nonbiased BLG-FET. In 
contrast, hole doping is so enhanced for the BLG-FET of $\Vgad=80$~V
that it took only 4 min, on the other hand so suppressed for that of
$\Vgad=-50$~V that it took more than 1000 min. The doping density 
increases almost linearly with respect to $\log t$ 
for $\Vgad=+80$~V and $+40$~V, whereas 
superlinearly for $\Vgad=0$~V and $-50$~V. The plots for runs~2 and 3,
having common $\Vgad = +80$~V,
are completely on the same line, which verifies that the thermal 
annealing in vacuum for the reproducing of the undoped state in the BLG 
does not affect the behavior of adsorption.

\begin{figure}
	\centering
	\includegraphics[clip]{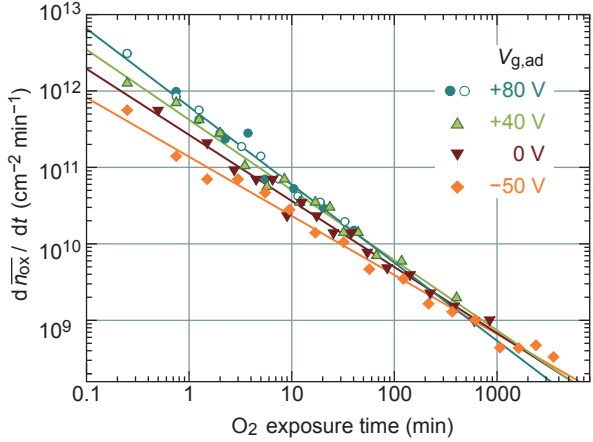}
	\caption{Double logarithmic plot of time dependence of $\tdiff{\nimp}{t}$. 
	Symbols are taken in common with Figure 2. We estimate
	$\tdiff{\nimp}{t}$ from the differential between the neighboring data 
	points for each run in Figure~\ref{f:timeevolve}. Lines are linear fits for all the 
	differential data. The slope of each curve gives $u=$ 1.02, 0.91, 0.86, 
	and 0.77 ($\tdiff{\nimp}{t}\propto t^{-u}$) for the gate voltage
	$\Vgad =+80$, $+40$, 0, and $-50$~V, respectively.}
	\label{f:rate}
\end{figure}

Figure~\ref{f:rate} shows the time dependence of the doping rate
$\tdiff{\nimp}{t}$ estimated from the differential between the 
neighboring data points in Figure~\ref{f:timeevolve}.
It is obvious that the doping rate changes in 
accordance with $\tdiff{\nimp}{t}\propto t^{-u}$. The power $u$ is 
dependent on $\Vgad$; $u\approx 1$ for $\Vgad =+80$~V and 
it decreases for the runs with lower $\Vgad $. This deviates from 
the conventional Langmurian kinetics for molecular adsorption which 
would give $\tdiff{\nimp}{t}\propto \exp(-t/\tau)$ with a constant $
\tau $. 

\begin{figure}[!t]
	\centering
	\includegraphics[clip]{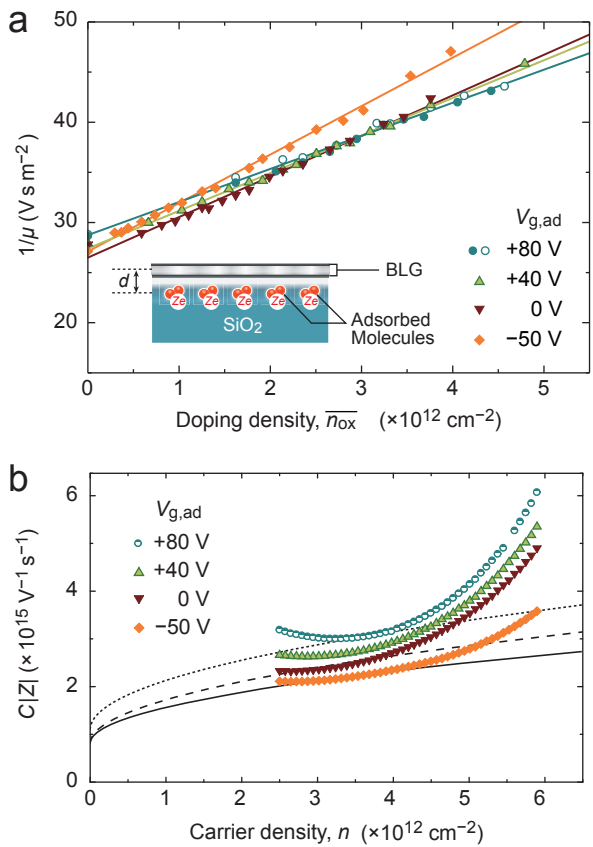}
	\caption{(a) Inverse mobility $\mu ^{-1}$~vs the doping density
	$\nimp$ for each run in Figure 2. Symbols are taken in common 
	with Figure~\ref{f:timeevolve}. Lines are linear fits (as for $\Vgad =+80$~V, the 
	data for both run~2 and run~3 are included). Inset: Adsorbed oxygen 
	species with charge $Ze$ positioned at the distance $d$ away from 
	the center of BLG. (b) $C\vert Z\vert $~vs the carrier density 
	$n$ (see Eq.~\eqref{eq:mobility}). 
For $\Vgad=+80$~V, $C\vert Z\vert $
	 is acquired by gathering the data for run~2 and run~3. Theoretical 
	results based on a charged-impurity scattering mechanism are shown 
	with the various distance $d$ and charge/molecular ratio $Z$:
	the solid line represents the result for $d=0.43$~nm and $Z=0.38$,
	the dashed line for $d=0.73$~nm and $Z=0.38$,
	and the dotted line for  $d=0.43$~nm and $Z=0.28$,
	respectively.}
	\label{f:mobility}
\end{figure}

Careful verification is necessary to inquire the gate-voltage-dependent 
and non-Langmurian temporal change of the molecular doping since the 
rate for doping density $\tdiff{\nimp}{t}$ is related to both of the 
rate for the chemisorption of molecules ($\tdiff{\Nox}{t}$, where $
\Nox$ is the areal density of the adsorbed oxygen molecules) and the 
transferred charge per adsorbed molecule (the charge/molecular ratio, $
Z$). Therefore, we analyze the mobility that includes the information 
of the scattering mechanism of the conducting electrons and the charge 
of the adsorbed molecules. Within a standard Boltzmann approach\cite{DasSarmaS2011}, the 
mobility is changed inversely proportional to the density of the 
scattering centers (i.e., the adsorbed molecules), $\Nox$. In the 
realistic case, the inverse mobility is given as a function of $\Nox$
and the carrier density $n$, which reads\cite{ChenJH2007}
\begin{equation}
	\frac{1}{\mu (n,\Nox)}=\frac{\Nox}{C(n)}+\frac{1}{\mu _{0}(n)}
	\label{eq:mobility}
\end{equation}
Here $\mu _{0}(n)$ represents the mobility of the pristine graphene 
without the adsorbed oxygen. The coefficient $C(n)$
represents the feature of carrier scattering by the adsorbed oxygen. 
On the one hand, the charged-impurity scattering\cite{AdamS2008}
gives $C(n)\propto {\left[1+6.53\sqrt{n}\left(d+\lambda _\mathrm{TF}\right)\right] }/{Z^{2}}$
for the BLG in the low-carrier-density regime and in the 
limit of $d\to 0$ within the Thomas--Fermi approximation\cite{AdamS2008},
where $d$ is the distance between the impurities 
and  the center of the two layers of the BLG (see the 
inset of Figure~\ref{f:mobility}a for the definition), and the screening length $\lambda 
_\mathrm{TF}={\kappa \hbar^{2}}/{4m^{*}e}\approx 1$ nm
($\kappa$: dielectric constant)\cite{AdamS2008}.
On the other hand, the short-range delta-correlated scatterers give the constant
$C(n)\equiv C_\mathrm{s}$\cite{AdamS2008}, or the strong impurities with the potential radius $
R$ give $C(n)\propto\lbrack\ln(R\sqrt{\pi n})\rbrack^2$ in the off-resonance condition
\cite{MonteverdeM2010}, which is a decreasing function of $n$ in the regime of
$n\sim 10^{12}\ \mathrm{cm^{-2}}$ taking $R$ to be several angstroms. As for the 
relation between the concentration of the adsorbed oxygen
molecules $\Nox$ and the $\mathrm{O_2}$-induced doping 
density $\nimp$, we assume that the charge $Ze$ of each 
adsorbed oxygen molecule dopes the carriers $-Ze$ in the BLG, notably,
$\nimp=-Z\Nox$. To be exact, the amount of the induced charge 
is not such a simple function\cite{XiaoS2009} 
proportional to the number of adsorbed molecules due to 
the energy-dependent DOS of BLG
\cite{McCannE2006, AdamS2010}
and the anomalistic screening effect therein
\cite{AdamS2007,ChenJH2007,XiaoS2009}. Yet in the low energy regime of 
BLG where the DOS is envisaged to be constant, the assumption above is 
appropriate.

Figure~\ref{f:mobility}a shows the inverse Drude mobility $\mu^{-1}$~vs
$\nimp$ plots at the carrier 
density of $n=2.5\times\mathrm{10^{12}\ cm^{-2}}$ (marked by the open circles 
in Figure~\ref{f:intro}b for run~1:$\Vgad=0$~V) for $\Vgad=+80$,
$+40$, $0$, and $-50$~V. Linear increase in $\mu ^{-1}$ with respect to
$\nimp$ was found. This, along with the 
linearity between $\mu ^{-1}$ and $\Nox$ given by Eq.~\eqref{eq:mobility}, 
implies that $Z$ is not a function of $\Nox$, i.e.,
invariant against the increase of the adsorbed 
molecules. Interestingly, the slope of $\mu ^{-1}$~vs $\nimp$ 
plot (the \textit{inverse} of the slope corresponds to $C(n)\vert 
Z\vert$ in Eq.~\eqref{eq:mobility}) depends on $\Vgad$. 
Note that before $\mathrm{O_2}$ was introduced ($
\nimp=0$), we observed $\mu ^{-1}\approx 28~\mathrm{V\, s\, m^{-2}}
$ irrespective of $\Vgad$, and thus the difference in the mobility 
by $\Vgad$ genuinely results from the adsorbed oxygen instead of 
other unintentional impurities on the BLG or the $\mathrm{SiO_2}$ substrate. 
In Figure~\ref{f:mobility}b, the inverse of the slope, $C(n)\vert Z\vert$, is shown for 
the various carrier densities, $n$. Therein we omit the data in the low 
carrier regime of $n<2.5\times 10^{12}\  \mathrm{cm^{-2}}$, in which the 
residual carriers due to electron--hole puddles cannot be disregarded and 
the carrier density $n$ (and thus also the Drude mobility) cannot be 
correctly estimated only by considering the gate electric field effect\cite{TanYW2007}. 
Similarly to the charged impurity model rather than otherwise, $
C(n)\vert Z\vert$ is increasing with $n$. The dependence experimentally 
observed, however, still deviates from the theoretical 
calculated results within the charged impurity model
plotted in Figure~\ref{f:mobility}b for 
various $d$ and $Z$ (assuming $d$ and $Z$ are invariant to $n$).

The difference in $C(n)\vert Z\vert $ depending upon $\Vgad$ indicates 
that the electronic polarity of graphene varies the adsorption states
of oxygen molecules, leading to the variation of $d$ and $Z$.
When the positive (negative) $\Vgad$ is applied,
negative (positive) carrier is electrically induced on graphene,
which may modify the interaction between graphene and 
the adsorbed oxygen molecules with the negative charge,
e.g., the Coulomb interaction and the overlap of the orbitals.
Eventually, the stable adsorption state is varied by $\Vgad$,
leading to the difference in mobility.
Besides, let us recall that conductivity measurement process
is set apart from the $\mathrm{O_2}$ adsorption process and that 
constant $\Vgad$ is not applied
when the mobility is measured (Figure~\ref{f:intro}a).
Accordingly, whereas the stable adsorption state of
oxygen during the conductivity measurement may 
differ from that during adsorption, 
the adsorbed oxygen molecules are kept in the former state 
during conductivity measurement, 
and the mobility varying by $\Vgad$ is actually observed.
This indicates
that the energetic barrier exists for charge redistribution 
between graphene and the adsorbed oxygen molecules (shown below),
and once the adsorption is accomplished, 
the charge $Ze$ on each adsorbed oxygen molecules 
will not immediately change just after switching on/off the gate voltage.
One possible reason
for the deviation between the experimental results 
and theoretical curve is that $d$ varies accompanied
with the change in $n$ (or sweeping $V_\mathrm{g}$).
Yet actually, 
since modifying $d$ by several angstroms results in the small change 
in $(C(n)\vert Z\vert )$ as shown in Figure~\ref{f:mobility}b, 
it is necessary to investigate more about 
the behavior of the adsorbed oxygen molecules in the gate electric field
in a the future study.

It is controversial what kind of oxygen species does actually cause the 
hole doping to graphene. Because the electron affinity of $\mathrm{O_2}$
($0.44$~eV\cite{ErvinKM2003}) is much lower than the work function of graphene (4.6~eV\cite{SqueSJ2007}), 
direct charge transfer between them seems unfavorable. Instead, 
with an analogy of the charge doping of diamond surface\cite{ChakrapaniV2007}, there is 
a widely accepted\cite{LevesquePL2010, PintoH2010, RyuS2010} hypothesis that the hole doping 
proceeds through an electrochemical reaction\cite{ChakrapaniV2007}
such as: $\mathrm{O_2}+2\mathrm{H_2O}+4e^{-}=4\mathrm{OH^-}$, by which the 
charge transfer is favorable due to the lowered free energy change 
$\Delta G=-0.7$~eV\cite{PintoH2010} on the condition that the oxygen pressure 
is 1 atm and $\mathrm{pH}=7$. 
This
electrochemical reaction needs the aid of 
water that is mostly eliminated in the experiment by annealing
(we observed no hysteresis in the $\sigma$ vs $V_\mathrm{g}$ curve,
that is, there are few charge traps often attributed to residual moisture
 on graphene or its substrate).
Yet as for 
graphene deposited on the hydrophilic $\mathrm{SiO_2}$ substrate\cite{AguirreCM2009}, 
it is possible that a small amount of residual water molecules (more than the chemical equivalent of 
$\mathrm{O_2}$) are trapped on $\mathrm{SiO_2}$ surface or voids, which cannot be 
easily removed by the vacuum annealing at 200~$^{\circ}\mathrm{C}$
in comparison with those on graphene surface. We suggest that 
the electrochemical mechanism is plausible also in our case, yet the 
adsorbed molecule could be other chemical species than $\mathrm{OH^-}$, the 
charge of which may be dependent on $\Vgad$.

\begin{figure}
	\centering
	\includegraphics[clip]{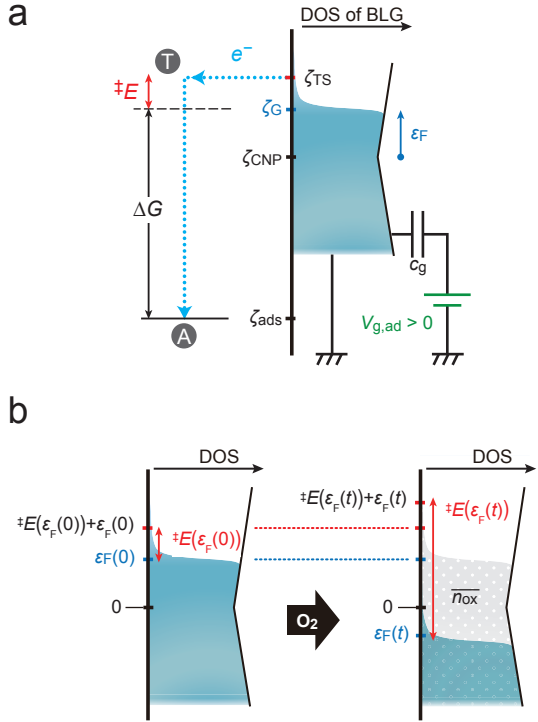}
	\caption{Schematic energy diagrams of the kinetics of $\mathrm{O_2}$ adsorption 
	(\Hkin). (a) Path for electron transfer in this model is shown 
	by the blue dotted arrow; electrons in BLG (the electrochemical 
	potential $\zeta _\mathrm{G}$) are transferred to $\mathrm{O_2}$ molecules via 
	the transition state (the circled \,T at the level of $\zeta _\mathrm{TS}$
	), giving the adsorbed oxygen species (the circled \,A at the level of 
	$\zeta _\mathrm{ads}$). The activation energy, the free energy change, and 
	the level of the CNP are denoted by ${}^{\ddagger }E$,
	$\Delta G$ and $\zeta _\mathrm{CNP}$, respectively. The Fermi level is defined by
	$\feref =\zeta _\mathrm{G}-\zeta _\mathrm{CNP}$. As a demonstration, the case 
	for $\Vgad >0$ is presented. (b) Temporal change in the activation 
	energy and the Fermi energy due to the adsorption of oxygen molecules to 
	BLG negatively doped by the positive gate voltage (as in case of  
	panel a). The left panel represents the case before oxygen adsorption 
	($t=0$), and the right panel represents the oxygen exposure for the 
	time $t$. Here the energy is measured from the CNP. The red and the 
	blue tick marks denote the level for the transition state and the Fermi 
	level, respectively. Oxygen adsorption lowers the Fermi 
	level, accompanied with the increase in hole doping of $\nimp$ (equal to 
	the area of grayed part, $\int_{\feref (t)}^{\feref(0)}{D(\feref )
	\>\mathrm{d}\feref }$), and the activation energy 
	increases according to Eq.~\eqref{eq:Butler}}
	\label{f:model}
\end{figure}

In light of 
the discussion above, the electrochemical
description\cite{HellerI2006, LevesquePL2010}
is expected to be applicable for the observed adsorption 
kinetics of oxygen to the BLG. Here we premise that the molecular 
adsorption is determined by the electrochemical potential of graphene, 
and consider the charge transfer kinetics in an approach based on 
Butler--Volmer theory\cite{BockrisJ1973,BockrisJ2000}. 
The model is 
schematically depicted in Figure~\ref{f:model}. The probability of the adsorption 
reaction is determined by the electrochemical potential of graphene ($
\zeta _\mathrm{G}$) and that in the equilibrium condition of the 
oxygen-chemisorption reaction ($\zeta _\mathrm{ads}$). 
When
$\Delta G=\zeta _\mathrm{ads}-\zeta _\mathrm{G}<0$,
the electrons favorably transfer from of graphene 
to the adsorbed oxygen (denoted by ``\,A'' in Figure~\ref{f:model}a),
and the oxygen-adsorption reaction proceeds. 
For charge transfer, 
the electrons should go through some energy
barrier; we assume that electrons tunnel from BLG to the $\mathrm{O_2}$ 
molecules via a \textit{single} transition state (denoted by ``\,T''), 
whose electrochemical potential is $\zeta _\mathrm{TS}$. 
The difference ${}^{\ddagger}E=\zeta _\mathrm{TS}-\zeta _\mathrm{G}$
corresponds to the activation 
energy of the oxygen-chemisorption reaction, which determines the 
frequency of the electron transfer. Whereas $\zeta _\mathrm{G}$ is dependent 
on the Fermi level $\feref$ as $\zeta _\mathrm{G}=\feref+\zeta 
_\mathrm{CNP}$ ($\zeta _\mathrm{CNP}$ is the electrochemical potential of the 
CNP), we envisage that $\zeta _\mathrm{TS}$ (or ${}^{\ddagger }E$) is a 
function of $\feref $ as well. 
In the framework of the Butler--Volmer 
theory, we obtain the dependence of ${}^{\ddagger }E$ on $\feref$ as
\begin{equation}
	{}^{\ddagger }E(\varepsilon _\textrm{F}+\mathrm{d}\feref )
		={}^{\ddagger }E(\feref )-\alpha\, \mathrm{d} \feref,
		\label{eq:Butler}
\end{equation}
where $\alpha (>0)$ is a constant related to the 
``transfer coefficient'' in the Butler--Volmer theory 
that associates the activation energy with \textit{the electrochemical potential}
(not the Fermi energy); thus herein we call $\alpha$ as ``pseudo transfer coefficient''
(see Supporting Information for detail in the derivation).
That is, we 
have the assumption that the activation energy scales linearly with the 
Fermi energy. Further assuming that the molecular adsorption rate $
\tdiff{\Nox}{t}$ is controlled by the electron transfer process
and is not strongly affected by other contributions such as molecular diffusion\cite{RomeroHE2009},
it is given by
\begin{equation}
	\diff{\Nox}{t}=\chi D\left(\feref +{}^{\ddagger }
		E(\feref )\right) f\left(\feref +{}^{\ddagger }E(\feref )
		;\feref \right),
		\label{eq:tunnel}
\end{equation}
where $D(\varepsilon )$ is the density of states (DOS) of BLG
($\varepsilon $ is the energy measured from $\zeta _\mathrm{CNP}$) and $f(\varepsilon ;
\feref)=\lbrack 1+{(\varepsilon -\feref )}/{\kbt}\rbrack^{-1}$
is the Fermi--Dirac distribution function. The coefficient $\chi $ 
does not depend on $\feref $
(if the distance between the adsorbing molecules and BLG varied 
depending on $\feref$ or $\Vgad$, the tunneling frequency 
would be affected so that $\chi$ might be dependent on them as well;
yet herein we ignore such effect for simplicity).
The right-side of Eq.~\eqref{eq:tunnel} 
represents the tunneling rate of the electron from the graphene to 
oxygen at the energy level of transition state,
$\varepsilon =\feref+{}^{\ddagger }E(\feref )=\zeta _\mathrm{TS}-\zeta _\mathrm{CNP}$.
According to Eq.~\eqref{eq:Butler} and Eq.~\eqref{eq:tunnel},
the molecular adsorption rate is dependent on the Fermi 
level of graphene. Using them, we can explain both the temporal 
evolution of the doping rate and its dependence on the gate voltage $\Vgad $.

Let us discuss the temporal change in the doping rate. 
The Fermi level 
of graphene is lowered with the increase of the adsorbed oxygen molecules, 
because the positive charge is induced on graphene by the charge $Ze$ 
they possess. 
Recalling that $\nimp=-Z\Nox$, the doping 
rate is given by $\tdiff{\nimp}{t}$. Furthermore, since
${}^{\ddagger }E \gg \kbt$ is fulfilled as shown later, we also 
approximate that $f(\feref +{}^{\ddagger }E(\feref );
\feref)\simeq \exp(-{}^{\ddagger }E(\feref )/\kbt)$. Using the relation $
\mathrm{d}\nimp=-D(\feref )\mathrm{d}\feref $, we acquire a formula 
describing the temporal change of the Fermi level:
\begin{widetext}
\begin{equation}
	-D\left(\feref (t)\right)\diff{ \feref (t)}{t}
	=\frac{p\kbt}{\alpha _\mathrm{te}}D\left(\feref (t)
	+{}^{\ddagger }E\left(\feref (t)\right)\right)
	\exp\left(\alpha _\mathrm{te}\frac{\feref (t)
	-\feref(0)}{\kbt}\right),
	\label{eq:diffeq}
\end{equation}
\end{widetext}
where $\feref(t)=\feref(t,\Vgad)$, expressing that the Fermi level is a
function of the exposure time $t$ and the gate voltage $\Vgad$,
and $\feref(0)=\feref(0,\Vgad)$, the Fermi level at $t=0$.
We have specifically defined two constants, 
the pseudo transfer coefficient $\alpha_\mathrm{te}$
(the subscript ``te'' abbreviates ``temporal evolution'') and
\begin{equation}
	p=-\frac{\alpha_\mathrm{te}Z\chi}{\kbt}\exp
		\left(-\frac{{}^{\ddagger }E\left(\feref (0)\right)}{\kbt}\right)
	\label{eq:pdef}
\end{equation}
The right side of Eq.~\eqref{eq:diffeq}
represents the product of the charge transfer frequency and the amount of charge per adsorbed molecule,  
whereas the left side does the resultant amount of the doped charge. 
Because BLG (or also single layer graphene) has the low DOS around the 
CNP compared to metal, the small amount of carrier doping results in the 
large shift in the Fermi level, which effectively controls the kinetics. 
Thus the adsorption kinetics is well described by  Eq.~\eqref{eq:diffeq}, the equation 
focusing on the Fermi level. 
When we envisage that the BLG is approximately described 
by two-dimensional parabolic dispersion of the free electron, the DOS 
becomes constant as $D\equiv D_\mathrm{P}=
{\gamma _{\perp }}/{\pi (\hbar v_\mathrm{F})^{2}}$, where
$v_\mathrm{F}=({\sqrt{3}}/{2})\gamma _{0}a/\hbar$ is the Fermi velocity in SLG, 
$a=2.46$ \AA\ is the in-plane lattice constant, and $\gamma _{0}=3.16$~eV and $
\gamma _{\perp }\approx 0.4$~eV\cite{ToyWW1977, DresselhausMS2002} 
are the intrasheet and intersheet transfer 
integrals, respectively. In this case (hereafter labeled as 
\Pkin ), Eq.~\eqref{eq:solPkin} is readily integrated, giving
\begin{equation}
	\nimp (t)=-\lbrace \feref (t)-\feref (0)\rbrace D_\mathrm{P}
	=\left(\frac{\kbt D_\mathrm{P}}{\alpha _\mathrm{te}}\right)\ln (1+pt)
	\label{eq:solPkin}
\end{equation}
Note that Eq.~\eqref{eq:solPkin} satisfies $
\tdiff{\nimp}{t}\propto \exp\lbrack -{\alpha 
_\mathrm{te}\nimp(t)}/{\kbt D}\rbrack$ and is equivalent to the integrated 
form of Elovich equation\cite{AharoniC1970, McLintockIS1967}, the empirical equation that is 
widely applicable to chemisorptions onto semiconductors. When the 
hyperbolic DOS of BLG\cite{AdamS2010} is reflected to Eq.~\eqref{eq:diffeq}, more accurate but 
more complicated expression of $\feref(t)$ is acquired (denoted 
as \Hkin ), given by
\begin{equation}
		-\frac{\alpha_\mathrm{te}}{\kbt}\exp\left(
		\alpha_\mathrm{te}\frac{\feref (0)}{\kbt}\right)S[\feref(0),\feref(t)] + pt =0,
	\label{eq:solHkin}
\end{equation}
where $S[\feref(0),\feref(t)]$ is a function that depends on the DOS 
at the Fermi level and that at the transition state level
(derived in the Supporting Information).

We performed curve fitting of the experimental results of $\nimp(t)$ 
 with Eq.~\eqref{eq:solPkin} and Eq.~\eqref{eq:solHkin} for \Pkin\ and \Hkin, respectively. 
The difference in $\Vgad$ by runs is simulated by the dependence of 
$\feref(0)$ on $\Vgad$ 
first without considering the gap-opening effect\cite{OhtaT2006, CastroEV2007}, 
that is, we calculate $\feref(0)$ using the relation that the charge 
of $c_\mathrm{g}(\Vgad-\Vcnp^0)$ doped on BLG by applying 
$\Vgad$ is equal to $\int_0^{\feref (0)} D(\varepsilon) \mathrm{d}\varepsilon$.
Irrespective of the kinetic models,
All the theoretical curves are well fitted to the experimental results 
except those for run~4 in the range above $10^{3}$ min. 
Eq.~\eqref{eq:diffeq} is invalid in the first place in the long-time regime in which 
the adsorption rate is almost as low as the desorption rate, since the 
present treatment includes no contribution of desorption. The deviation 
is, however, contrary to the expectation; the desorption should suppress 
the evolution of the hole doping yet the enhanced doping was actually 
observed. Thus we suspect that it is due to long-time scale 
chemisorption of oxygen onto graphene which is ubiquitously observed in 
carbon materials\cite{KobayashiN1998, SumanasekeraGU2010}.
Note that the several volts of offsets are 
added in $\Vshift$ (corresponding to the doping density of $
2\times 10^{11}\ \mathrm{cm^{-2}}$) to improve the fitting in the time range of 
$t\le 10^{0}\ \mathrm{min}$. This small offset corresponds to a fast reaction 
that finishes at the very initial stage of adsorption, e.g., due to the 
reactive chemisorption of $\mathrm{O_2}$ to the defect or edge site of 
graphene\cite{WangX2010, KoehlerFM2010}.

\begin{figure}[!t]
	\centering
	\includegraphics[clip]{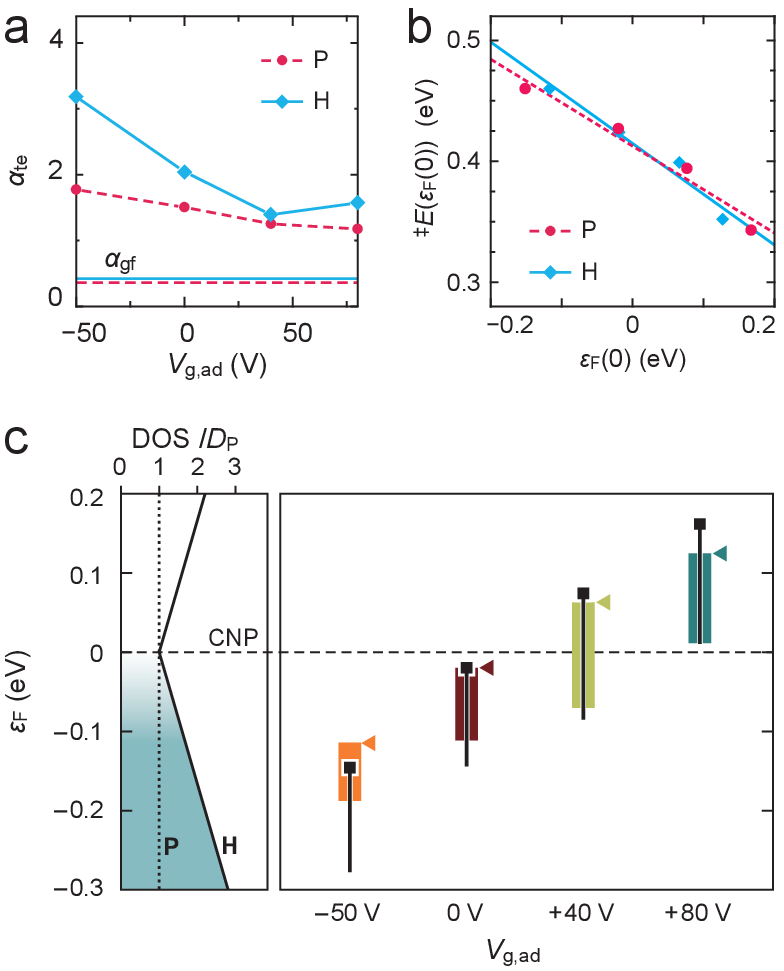}
	\caption{Summary of parameters obtained by the curve fitting in Figure~\ref{f:timeevolve} 
	with \Pkin\ (P) and 
	\Hkin\ (H) compared. (a) Dependence of the pseudo transfer coefficient 
	$\alpha_\mathrm{te}$ on $\Vgad $. The 
	pseudo transfer coefficient $\alpha _\mathrm{gf}$ acquired by the linear fitting to 
	panel b is also plotted. (b) Initial activation energy
	${}^{\ddagger }E(\feref (0))$ plotted as a function of 
	$\feref (0)=\feref (0,V_\mathrm{g})$.
	(c) (Left panel) Density of states (DOS, solid line) for BLG as a function of the
	Fermi energy, where $D_\mathrm{P}$ (DOS at the energy of $\feref=0$)
	is taken as unit of the DOS.
	The solid line denotes the DOS applied for the \Hkin .
	In \Pkin\ the DOS is treated as constant; $D=D_\mathrm{P}$ (dotted line).
	(Right panel) Change of the Fermi 
	level by oxygen adsorption as a function of $\Vgad $. Bars and 
	boxes represent the range calculated under \Pkin\ and \Hkin, 
	respectively, assuming $\gamma _{\perp }=0.4$~eV. 
	The Fermi level before the oxygen adsorption ($\feref(0)$)
	is at the point indicated by 
	triangles (black squares) for H (P) kinetics,
	whereas the Fermi level shifts downward with the increase of
	the adsorbed oxygen, and eventually reaches the bottom end of the boxes
	(bars) in the end of the oxygen adsorption for the H (P) kinetics. 
	The energy is measured from the CNP (dashed line), and the 
	gap-opening effect is not considered. }
	\label{f:fits}
\end{figure}

Figure~\ref{f:fits} shows the profiles of the fitting results. 
Practically,
the fitting parameters
are following two: $\alpha _\mathrm{te}$ and $p,$ but in 
order to derive the initial activation energy before the oxygen adsorption,
${}^{\ddagger }E(\feref(0))$, from $p$ based on Eq.~\eqref{eq:solPkin}, 
we briefly assume that the charge/molecular ratio is independent 
of the gate voltage and determine that $Z\chi 
\gamma _{\perp }=1\times 10^{5}\ \mathrm{eV^{2}\min ^{-1}}$
from the pre-exponential factor in the literature\cite{LevesquePL2010}, 
${c_\mathrm{ox}\nu \kappa _\mathrm{el}}\sim
{10^{17}\ \mathrm{cm^{-2}s^{-1}}}$
($c_\mathrm{ox}\nu \kappa _\mathrm{el}$ in the literature 
corresponds to $\chi \gamma _{\perp }/\pi (\hbar v_\mathrm{F})^{2}$ in this paper).
Note that if we assume 
another value than $Z\chi \gamma _{\perp }
=1\times 10^{5}\ \mathrm{eV^{2}\min ^{-1}}$,
 it results in a uniform shift of the calculated $
{}^{\ddagger }E(\feref (0))$. In addition,
the charge/molecular ratio 
of the adsorbed molecule $Z$ is likely dependent on $\Vgad $ from 
the discussion about the mobility, but it leads the shift of 
$^{\ddagger }E(\feref (0))$ only by $\sim \kbt=0.026$~eV.
On the one hand,
$\alpha _\mathrm{te}$ (representing the temporal change of the activation energy)
exhibits a significant 
deviation between the \Hkin\ and the \Pkin (Figure~\ref{f:fits}a), or depends 
on the treatment of the DOS of BLG. This deviation is understood as 
follows: \Hkin\ reflects the DOS of BLG that is a 
monotonically increasing function with respect to $\vert \feref\vert $
having a minimum ($D=D_\mathrm{P}$) at the CNP (Figure~\ref{f:fits}(c)), yet 
\Pkin\ does not. Since the downward shift of the Fermi level upon 
the oxygen adsorption depends on the DOS at the Fermi level, the range 
of the change in the Fermi level differs between two kinetics models (Figure~\ref{f:fits}(c)). 
Thus $\alpha _\mathrm{te}$, 
related with the Fermi level by Eq.~\eqref{eq:Butler}, is calculated differently.
This result is contrasted
with the conventional electrochemical reaction on the metal electrodes 
in which the kinetics is not significantly affected by the DOS of the 
electrodes and is just owing to the low DOS of the BLG with comparison 
to the metal. 
On the other hand, we found that 
${}^{\ddagger }E(\feref (0))$ is a decreasing function 
of $\feref(0)=\feref (0, \Vgad )$
(shown in Figure~\ref{f:fits}b, therein we plot ${}^{\ddagger }E(\feref (0))$
with respect to $\feref(0)$ instead of $\Vgad$).
From Eq.~\eqref{eq:Butler}, the slope of the plots in Figure~\ref{f:fits}b
corresponds to the pseudo transfer coefficient $\alpha $, and we found 
that $\alpha _\mathrm{gf}=$ 0.36 and 0.42 for \Pkin\ and \Hkin, 
respectively (the subscript ``gf'' abbreviates ``gate electric field''). 
Herein we distinguish $\alpha _\mathrm{gf}$ from $\alpha _\mathrm{te}$; being 
different from $\alpha _\mathrm{te}$ that is calculated based on the temporal 
Fermi level shift \textit{by oxygen adsorption} (thus $\alpha _\mathrm{te}$ 
inevitably includes the oxygen adsorption effect), $\alpha _\mathrm{gf}$ is 
acquired by tuning the Fermi level \textit{electrically} at $t=0$
before oxygen adsorption. 
We found that $\alpha _\mathrm{gf}$ is much smaller than 
$\alpha _\mathrm{te}(\ge 1)$,
or rather near 0.5, a typical transfer coefficient for simple 
redox reactions\cite{HellerI2006, BockrisJ1973} 
(Figure~\ref{f:fits}a).  %
Specifically, the oxygen adsorption effect included only in $\alpha 
_\mathrm{te}$ but not in $\alpha _\mathrm{gf}$ is attributed,
e.g., to the electric dipole layer\cite{GiovannettiG2008}
formed between graphene and the adsorbed molecules
and the Coulomb interaction between the adsorbed oxygen molecules.
We expect that these 
effects also raise the activation energy roughly in proportional to the 
number of molecules $\Nox$ (or the doping density $\nimp$), and thus 
we have the expression for the additional adsorption effect: $
\mathrm{d}^{\ddagger}E=\xi _\mathrm{mol}
\mathrm{d}\nimp=-\xi _\mathrm{mol}D(\feref)\mathrm{d}\feref$
($\xi _\mathrm{mol}$: the proportional coefficient;
herein we use the relation $\mathrm{d}\nimp=-D(\feref)d\feref$ again).
Then we acquire $\alpha _\mathrm{te}
\simeq\alpha _\mathrm{gf}+\xi _\mathrm{mol}\langle D(\feref)\rangle$
($\langle D(\feref)\rangle$ denotes the average DOS in the range of the Fermi level
for each run),
which indicates that $\alpha _\mathrm{te}
$ is large when the DOS is large in the level far away from the CNP (see 
Figure~\ref{f:fits}c). Indeed, as shown in Figure~\ref{f:fits}a,
$\alpha _\mathrm{te}$ for \Hkin\ shows the V-shaped 
dependence on $\Vgad$ with the minimum  for $
\Vgad=+40$~V, which gives the smallest average DOS. 
Though the kinetics for the molecular adsorption is affected by various 
effects as mentioned above, we represent them by the parameter $\alpha 
_\mathrm{te}$ and succeed in accounting for the observed kinetics in a facile 
way.

Finally let us account for the power-law dependence of $
\tdiff{\nimp}{t}\propto t^{-u}$ shown in Figure~\ref{f:rate}. 
It is helpful to look on the simpler \Pkin\ for 
the assessment of $\tdiff{\nimp}{t}$; from  
Eq.~\eqref{eq:solPkin}, we obtain $
\tdiff{\nimp}{t}\propto p/(1+pt)$.
Approximately
we have $\tdiff{\nimp}{t}\propto \left( u\langle t \rangle^{u-1}\right) t^{-u}$
where $u\simeq \lbrack 1+({1}/{p\langle t\rangle })\rbrack ^{-1}\le 1$ (the time 
$\langle t\rangle $ is the center of the expansion of $\ln 
(\tdiff{\nimp}{t})$ in terms of $\ln t$, and it is a good 
approximation if $p\langle t\rangle \gg 1$, or otherwise if $
p\langle t\rangle \simeq 1$ in the time range such that $t=\lbrack 
10^{-1}\langle t\rangle , 10\langle t\rangle \rbrack $).
Since $p$ is 
intensively dependent on $\Vgad$ (recall that 
(i)
$p$ is 
exponentially decaying against ${}^{\ddagger}E(\feref(0))$ as 
represented in Eq.~\eqref{eq:pdef}, 
(ii)
${}^{\ddagger}E(\feref(0))$ linearly 
decreases with respect to $\feref(0)$ 
with the slope of $-\alpha _\mathrm{gf}$ as shown in Figure~\ref{f:fits}c,
(iii)
 $\feref(0)$ is an increasing function of $\Vgad$.
We acquired $p=$19.6, 2.9, 0.98 and 0.32 for $\Vgad=+80$, $
+40$, $0$, and $-50$~V, respectively, by fitting within the
\Pkin\ model), we can find that $u$ is almost unity for a 
positively high $\Vgad$ and tends be smaller for negatively high $
\Vgad$ within the time range experimentally scoped, which is 
consistent with the observed behavior.
When the electrochemical mechanism governs the kinetics of the oxygen adsorption,
the activation energy of the charge transfer continually increases
 with the $\mathrm{O_2}$
exposure time increasing. This effect leads to non-Langmurian kinetics of the 
oxygen adsorption and the power-law decrease of $\tdiff{\nimp}{t}$,
even though neither the desorption process nor the saturation limit of adsorption
are taken into consideration.

In BLG, it is known that a band gap opens due to the energy difference between two 
layers\cite{CastroEV2010a}. The gate electric field as well as the adsorbed oxygen may 
produce such a strong energy difference that the eventual band gap 
should affect the time evolution of molecular adsorption. The band gap 
opening effect is expected to exhibit most prominently when the Fermi 
level goes across the CNP (at the point shown by arrows in 
Figure~\ref{f:timeevolve}), while we cannot find such behavior 
obviously. 
We guess it is partly because most of the adsorbed oxygen molecules exist
in the interface between graphene and the gate dielectric $\mathrm{SiO_2}$.
For the band gap opening effect to appear,
it is necessary that the gate electric field and the molecular field enhance 
each other when the Fermi level is near the CNP 
(i.e., the charge induced by the gate electric field and that by the adsorbed molecules
including unintentional residual impurities on the $\mathrm{SiO_2}$ substrate are balanced),
yet it is possible only when the molecules mainly adsorb on the top surface of BLG, 
and not for the molecules adsorbed in the interface (Figure~{S2}, Supporting Information).
Or it may be partly because the 
disordered potential due to the impurities in the substrate fluctuates 
the energy level  around which the band gap exists\cite{RutterGM2011}, 
eventually blurring band gap opening effects and chemical reactivity\cite{SharmaR2010} of BLG. Details about the band-gap opening effects are 
discussed in the Supporting Information.

In summary, we investigated the weak chemisorption of $\mathrm{O_2}$ 
molecules on bilayer graphene by measuring its transport properties. 
The hole doping due to $\mathrm{O_2}$
chemisorption is remarkably dependent on the gate voltage,
and the amount of the doped carrier increases with $\mathrm{O_2}$ exposure time, the 
rate of which is in accordance with $\propto t^{-u} (u\le 1)$ rather than with 
conventional Langmuirian kinetics. We conclude from these that an 
electrochemical reaction governs the $\mathrm{O_2}$ chemisorption process, 
in which the rate of the chemisorption is determined by the Fermi level 
of graphene, and indeed succeed in accounting for the observed kinetics by 
the analysis based on the Butler--Volmer theory. We also found that the 
chemisorbed molecules decrease the mobility of graphene, and 
interestingly, the mobility change is dependent on the gate voltage 
applied during the adsorption, indicating that the adsorption state, 
e.g., transferred charge or distance between a molecule and graphene, can 
be modified electrically. 
Graphene, offering a continuously tunable platform for study of chemisorptions on it,
realizes the electrical control of the adsorption by gate electric
field, a novel and versatile method in which we would explore extensively a
wider variety of host--guest interactions between graphene and foreign molecules.

\mbox{}\par
	\textbf{Supporting Information.}
	Additional descriptions about
	(i) the experimental method,
	(ii) the electrochemistry-based
	kinetics model and the mathematical derivation for it, and 
	(iii) an expanded discussion about the gap-opening effects.
\vspace{2em}

\begin{acknowledgements}
The authors acknowledge support from Grant-in-Aid for Scientific 
Research No. 20001006 from the Ministry of Education, Culture, Sports, 
Science and Technology, Japan.
The authors thank M. Kiguchi,  T. Kawakami, 
K. Yokota, and Y. Kudo  for useful discussion.
\end{acknowledgements}

% \begin{suppinfo}
% Additional descriptions about
% (i) the experimental method,
% (ii) the electrochemistry-based kinetics model and the mathematical derivation for it,
% (iii) the expanded discussion about the gap-opening effects are presented. 
% \end{suppinfo}

% \begin{tocentry}
% 	\centering
% 	\includegraphics{3DP4.eps}
% \end{tocentry}

% \bibliography{ref}

\end{document}